\documentclass[letter]{aa} 

\usepackage{url}
\usepackage{graphicx}
\usepackage{txfonts}
\usepackage{color}
\usepackage{natbib,twoopt}
\usepackage[hyphenbreaks]{breakurl}
\usepackage[breaklinks]{hyperref}      
\RequirePackage{siunitx}
\bibpunct{(}{)}{;}{a}{}{,}             
\definecolor{cobalt}{rgb}{0.06, 0.2, 0.65}
\hypersetup{
  colorlinks,
  citecolor=cobalt,
  linkcolor=[rgb]{0.8, 0.2, 1.0},
  urlcolor=cobalt,
}
\makeatletter
  \newcommandtwoopt{\citeads}[3][][]{\href{http://adsabs.harvard.edu/abs/#3}%
    {\def\hyper@linkstart##1##2{}%
     \let\hyper@linkend\@empty\citealp[#1][#2]{#3}}}
  \newcommandtwoopt{\citepads}[3][][]{\href{http://adsabs.harvard.edu/abs/#3}%
    {\def\hyper@linkstart##1##2{}%
     \let\hyper@linkend\@empty\citep[#1][#2]{#3}}}
  \newcommandtwoopt{\citetads}[3][][]{\href{http://adsabs.harvard.edu/abs/#3}%
    {\def\hyper@linkstart##1##2{}%
     \let\hyper@linkend\@empty\citet[#1][#2]{#3}}}
  \newcommandtwoopt{\citeyearads}[3][][]%
    {\href{http://adsabs.harvard.edu/abs/#3}
    {\def\hyper@linkstart##1##2{}%
     \let\hyper@linkend\@empty\citeyear[#1][#2]{#3}}}
\makeatother

\newcommand{\hi}{\mbox{H\,{\sc i}}}
\newcommand{\kms}{km\,s$^{-1}$}

\DeclareSIUnit \parsec {pc}
\DeclareSIUnit\jansky{Jy}

\begin{document}

   \title{Origin of gas in the Magellanic Bridge: MeerKAT detection of \hi\ 21-cm absorption}


   \titlerunning{Gas associated with the Magellanic bridge}

   \author{A. P. M. Morelli
          \inst{1}
          \and
          J. Kerp\inst{1}
          \and N. Gupta\inst{2}
          \and F. Combes\inst{3}
          \and S. A. Balashev\inst{4}
          \and P. Noterdaeme\inst{5}
          \and H. Chen\inst{6}
          \and K. L. Emig\inst{7}
          \and E. Momjian\inst{8}
          }

   \institute{Argelander-Institut für Astronomie, Auf dem Hügel 71, D-53121 Bonn, Germany
         \and
             Inter-University Centre for Astronomy and Astrophysics, Post Bag 4, Ganeshkhind, Pune 411 007, India
        \and
            Institut d’astrophysique de Paris, UMR 7095, CNRS, 98bis bd Arago, 75014 Paris, France
        \and 
            Ioffe Institute, 26 Politeknicheskaya st., St. Petersburg, 194021, Russia
        \and
            Institut d’astrophysique de Paris, UMR 7095, CNRS, 98bis bd Arago, 75014 Paris, France
        \and
            Departement of Astronomy and Astrophysics, The University of Chicago, Chicago, IL 60637, USA
        \and
            Observatoire de Paris, Coll\`ege de France, PSL University, Sorbonne University, CNRS, LUX, Paris, France
        \and
            National Radio Astronomy Observatory, 520 Edgemont Road, Charlottesville, VA 22903, USA
        \and
            National Radio Astronomy Observatory, 1011 Lopezville Road Socorro, NM 87801, USA
             }

   \date{Received April 3, 2025; accepted September 23, 2025}

 
  \abstract
   {} 
   {\hi\ 21-cm absorption lines are investigated to determine the origin of the neutral atomic hydrogen (\hi) of the Magellanic Bridge (MB).  Using the MeerKat Absorption Line Survey (MALS) data we report the detection of an \hi\ absorption line at a peak signal--to--noise ratio of 10 caused by MB gas against the radio source J033242.97-724904.5. In combination with earlier data obtained with the Australia Telescope Compact Array (ATCA) our new detected \hi\ line permits the exploration of the MB atomic hydrogen gas across 4-6\,kpc.}
   {The radial velocity profiles from the ATCA data and new data from MALS are analysed. Apart from the excitation conditions, the radial velocity structure of the \hi\ gas seen in emission and absorption is investigated. Eventually the gas-to-dust ratio is quantified to identify the origin of the MB gas being either from the SMC (Small Magellanic Cloud) or the LMC (Large Magellanic Cloud).}
   {The \hi\ absorption lines towards lines of sight separated by several kpc consistently coincide with the densest and perhaps coolest gas at the lower radial-velocity limit of the corresponding \hi\ emission profiles. The gas-to-dust ratio is found to be consistent with an origin of the MB gas from the LMC. The large scale velocity distribution as seen from the \hi\ absorption features favors the LMC-SMC direct collision scenario over the close fly-by scenario, as also currently found by numerical simulations.}
   {}

   \keywords{ISM: clouds --
                Magellanic Clouds --
                Radio Lines: galaxies
               }

   \maketitle
%

\section{Introduction}
The Magellanic Cloud System (MCS) is the closest extragalactic laboratory for studying structure formation on galactic scales. It also offers the rare opportunity in extragalactic science to measure the proper motions of galaxy pairs \citep[][and references therein]{Zivick2018}. Currently two models compete to explain the formation of the MB: (a) a close flyby \citep{Besla2012} or (b) a direct collision \citep[e.g.,][]{Besla2012, Belokurov2018, Zivick2018}.
While the gaseous morphology of observed structures may not be straightforward to interpret, the physical conditions of the neutral gas provide insights into its formation history.  
Here, we present a recently identified \hi\ 21-cm absorption line observed, part of the MeerKAT Absorption Line Survey \citep[MALS;][]{Gupta2016}, towards the Magellanic Bridge (MB), connecting in \hi\/ emission the LMC and the SMC. It complements information obtained by earlier investigations by \citet[][]{Kobulnicky1999}.
The MALS line of sight towards the quasar J033242.97-724904.5\footnote{Hereafter, referred as J0332-7249.} ($l$ = 288\fdg 589394, $b$ = -39\fdg 500745), presented here probes the MB gas away from the leading edge probed by B0312-770 and B0202-765 \citep{Kobulnicky1999}. 

Due to its very low rate of spontaneous emission, the \hi\ 21-cm line is a unique probe of the density structure and kinematics of the atomic ISM. Investigating the \hi\ 21-cm line simultaneously in emission and absorption reveals the excitation conditions as well as the relative fraction of cold and warm neutral atomic gas along a line of sight. The observations are carried out towards a luminous extragalactic background source, sufficiently bright to outshine the \hi\ line in emission. 

The population of the hyperfine ($F = 1$) upper level in the hydrogen atom relative to the ground state ($F=0$) is defined by the spin temperature $T_\mathrm{Spin}$ \citep[][]{Field1959}, which can be measured as

\begin{equation}
	T_{\rm Spin}  = \frac{N_{\mathrm{HI}}}{1.823\times10^{18}\int \tau \; \mathrm{d}v}\label{Eq:Tspin}
\end{equation}

Here, $N_\mathrm{HI}$ is in $\si{\centi\meter}^{-2}$, $T_{\mathrm{spin}}$ in $\si{\kelvin}$ and the velocity interval $\mathrm{d}v$  in $\si{\kilo\meter\second}^{-1}$. The measured optical depth $\tau$ of the absorption depends mainly on two quantities: the column density $N_\mathrm{HI}$ and the spin temperature $T_\mathrm{Spin}$, i.e., $\tau$  $\propto$ $\frac{N_\mathrm{HI}}{T_\mathrm{Spin}}$.
In the cold phase, the density maybe sufficient to drive through collisions $T_\mathrm{Spin}$ towards the kinetic temperature of the gas i.e., $T_{\rm Spin}$ $\sim$ $T_{\rm K}$ whereas in the warmer atomic gas, $T_{\rm Spin}$ $<$ $T_{\rm K}$ \citep[][]{Liszt01}. Generally, the quantity $T_{\rm Spin}$ derived from the observables represents the column density weighted mean of the spin temperatures along the line of sight \citep[e.g.,][]{KulkarniHeiles1988}.

The full width at half maximum (FWHM) of the line is related to the Doppler temperature ($T_\mathrm{D}$) of the gas as, 

	$T_{D}$ = $\frac{m_H\Delta v^2}{8\mathrm{k_B} \ln(2)}$  = $21.866\times\Delta v^2$[K],

with $\Delta v$ in units of $\mathrm{km\,s^{-1}}$  for hydrogen mass $m_{H}$ and Boltzmann's constant $\mathrm{k_B}$ \citep{Payne1980}. 
  
In most spectra, the line width is due to streaming and turbulent motions, broadening the line well beyond the thermal contribution,
and $T_\mathrm{D} >> T_{\rm K}$.
In particular, toward the Galactic Plane, the line of sight can probe blended multiple, unrelated physical structures.  
In the case of MCS, tidal forces tend to separate the different velocity components across the sky, reducing the complexity along the line of sight \citep{Bruens2005}.  

In our analysis, we evaluate \hi\ 21-cm absorption lines (resolution$\sim$10$^{\prime\prime}$) towards the MB from MALS and \hi\ 21-cm emission line information extracted from the HI4PI survey \citep[resolution $16\farcm4$;][]{HI4PI}.  

\citet[][]{Kobulnicky1999} have found $T_\mathrm{Spin}$ to vary appreciably (50 - 150\,K) across the MB. We adopt a distance of 50 kpc for the Magellanic System \citep{deGrijs2014}.

From the associated metallicity, the \hi\ 21-cm line will provide insights into the formation history of the MCS.

\section{Observations and Data Analysis}

\begin{figure}[!t]
	\centering
	\includegraphics[width=\hsize]{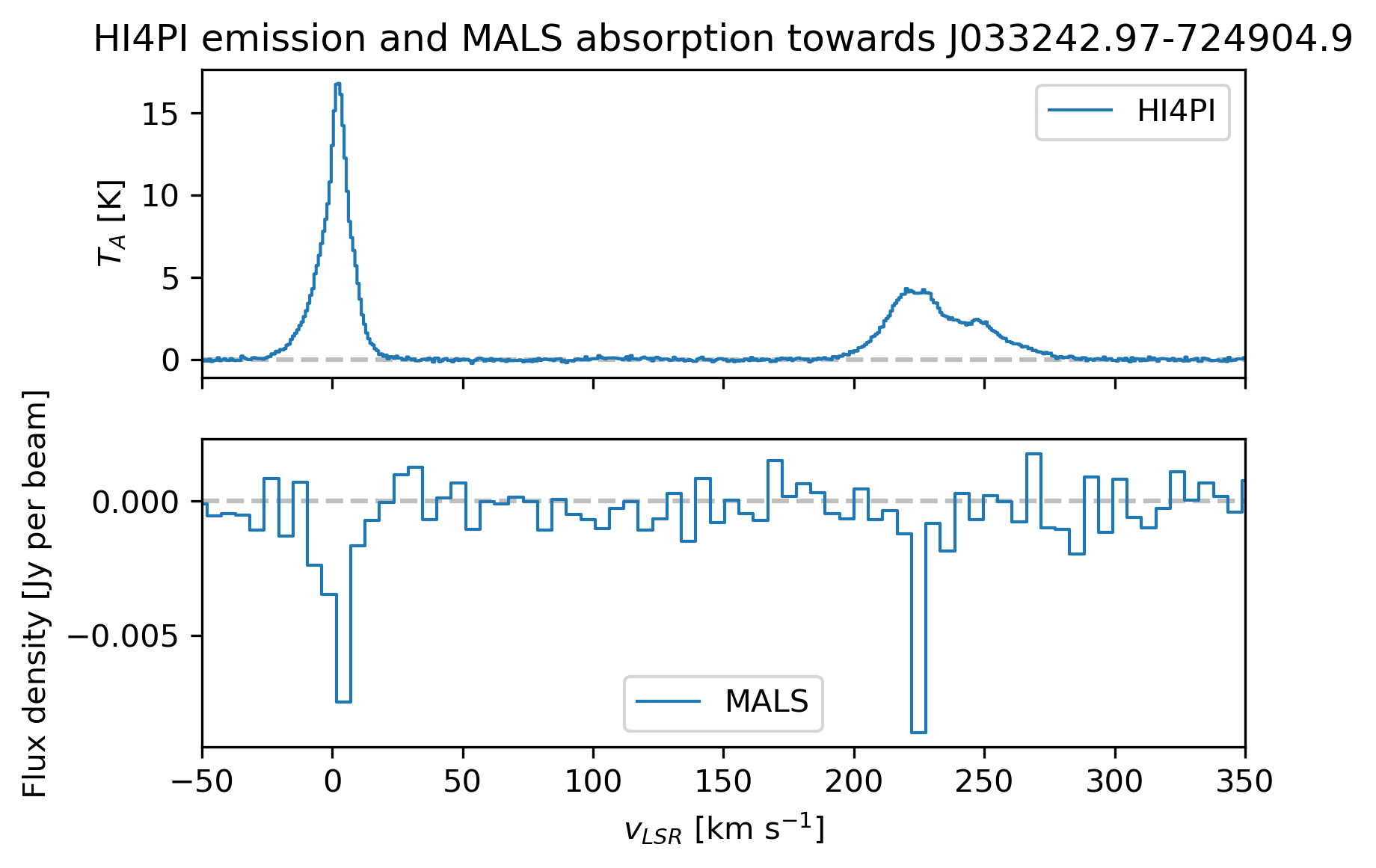}
	\caption{HI4PI emission and MALS absorption spectrum towards J0332-7249. The top panel shows the HI4PI \hi\ spectrum with $\Delta v_\mathrm{LSR}= 1.28\,\mathrm{km\,s^{-1}}$ velocity resolution. Single dish observations are sensitive to all \hi\ gas phases. The bottom panel shows the corresponding MALS absorption line spectrum with a coarser velocity resolution of $\Delta v_\mathrm{LSR}= 6.1\, \mathrm{km\,s^{-1}}$. Absorption line detection among the more dominant emission requires high angular resolution and is a tracer of cold and diffuse \hi\ gas. The Gaussian decomposition components of both \hi\ profiles are compiled in Tab.\,\ref{tab:emissionfits} and Tab.\,\ref{tab:absorptionfits} respectively.
    }
	\label{fig:Magellanic3}
\end{figure}

We observed the field centered at J0332-7249 using 60 antennas of the MeerKAT-64 array on 2021 January 18.  We used the 32K mode of the SKA Reconfigurable Application Board (SKARAB) correlator to split the total bandwidth of 856\,MHz centered at 1283.9869\,MHz into 32768 frequency channels. The resultant frequency resolution is 26.123\,kHz, or 5.5\,\kms, at the \hi\ 21-cm line frequency.  

The total on-source time on J0332-7249 was 56\,mins.
For the Galactic \hi\ 21-cm line analysis, we generated a measurement set comprising only XX and YY polarization products over 400 frequency channels centered at 1420.2709\,MHz in the {\it topocentric} frame. These data were processed using the Automated Radio Telescope Imaging Pipeline (ARTIP; Gupta et al. 2021) based on NRAO’s Common Astronomy Software Applications (CASA) package \citep{CASA2022}.
The details of Galactic spectral line processing and line search are provided in \cite{Gupta2025}.
The peak flux density of the unresolved continuum source is 244.1\,mJy\,beam$^{-1}$.  The unsmoothed spectrum,  extracted at the pixel corresponding to the peak flux density of the radio source, has an rms noise of 0.8\,mJy\,beam$^{-1}$.

\section{Results}

\begin{figure}
    \centering
    \includegraphics[width=\hsize]{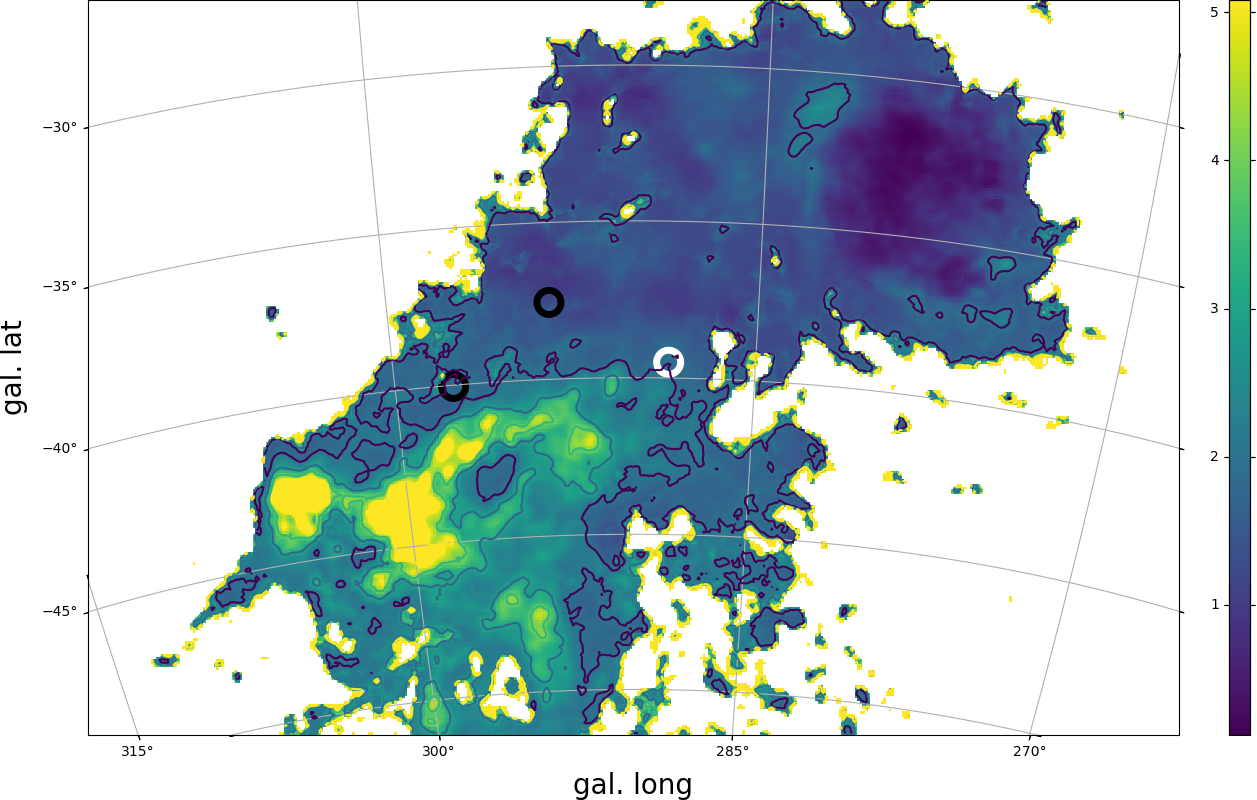}
    \caption{Gas--to--dust ratio of the MCS in units of the Milky Way's median value of $\frac{N_{\mathrm{H}}}{E\mathrm{(B-V)}}=(0.63\pm0.05)\times 10^{22}\frac{\si{\centi\meter}^{-2}}{\mathrm{mag}}$. Marked by contour lines are the 2, 3, 4 and 5 times higher values, for comparision with \cite{Welty2012} and \cite{Koornneef1982}. The circles mark the \hi\ absorption lines detected by \cite{Kobulnicky1999} (black), the white one marks the MALS pointing position reported here.}
    \label{fig:gastodustratios}
\end{figure}

The MALS \hi\ 21-cm absorption line spectrum of J0332-7249 along with the corresponding \hi\ emission line from HI4PI are presented in Fig.\,\ref{fig:Magellanic3}. \citet[][]{Kobulnicky1999} observed eight sight lines, five probing the outskirts of the MCS and three probing the MB, using the Australia Telescope Compact Array (ATCA). Their final ATCA spectral image cube had a high spectral resolution ($0.82\,\mathrm{km\,s^{-1}}$ channel separation).  The two sight lines, namely B0202-765 and B0312-770, probing MB towards the leading edge of the MCS, are detected in \hi\ 21-cm absorption (see Fig.\,\ref{fig:gastodustratios}).  
J0332-7249 probes a different location in the MB: it is half way between the SMC and the LMC, where the column density $N_\mathrm{HI}$ is twice compared to B0202-765 and B0312-770, with a higher mean radial velocity. The \hi\ emission line in all three cases shows two prominent peaks and the \hi\ absorption feature overlaps with the lower radial velocity component.  The ATCA absorption profile, possibly due to better spectral resolution, exhibits multiple narrow distinct \hi\ features.

Table\,\ref{tab:absorptionfits} compiles the Gaussian fit details of three Gaussian--components in the \hi\ emission profile towards J0332-7249 by \cite{KalberlaHaud2018}. Two of them are centered within a single spectra channel of $1.28\,\mathrm{km\,s^{-1}}$ width at $v_\mathrm{LSR} \simeq 219.1\,\mathrm{km\,s^{-1}}$ and $v_\mathrm{LSR} \simeq 219.6\,\mathrm{km\,s^{-1}}$ respectively. The third one is positioned at slightly higher radial velocity of $v_\mathrm{LSR} \simeq 228.3\,\mathrm{km\,s^{-1}}$. All three sum up to a column density of $N_\mathrm{HI}(J0332)= (1.18\pm0.07)\times 10^{20}\,\mathrm{cm^{-2}}$. Because we cannot determine by the emission line data the individual contribution of the emission profiles to the absorption feature observed, we apply a column density weighting to estimate the spin temperature. Using Eq.\,\ref{Eq:Tspin}, we calculate $T_\mathrm{Spin}(\mathrm{MB}) = 69 \pm 11$\,K. This is in between the spin temperature determined by \cite{Kobulnicky1999} for the lines of sight through the MB towards B0202-765 and B0312-770. 

Inherent to this analysis are the limitations due to the quantitative correlation of Parkes 64-m single dish \hi\ emission data and interferometric absorption line data with different spatial resolutions from arcsec to 16.2\,arcmin, definitely probing different physical gas volumes.  Accordingly, our estimated spin temperature for the MB is systematically biased to lower or higher values.

The coincidence between the lower velocity component in \hi\ emission and absorption implies that the atomic gas is denser at these velocities ($v_\mathrm{LSR} \simeq 219\, \mathrm{km\,s^{-1}}$).
Notably, the presence of two prominent velocity components in \hi\ emission towards the line of sight in Fig.~\ref{fig:Magellanic3} is not unique but a common feature throughout the whole MCS. It is also known that the gas with $v_\mathrm{LSR} \simeq 180\,\mathrm{km\,s^{-1}}$ has a lower radial velocity than that of the LMC at $v_\mathrm{LSR} \simeq 280\,\mathrm{km\,s^{-1}}$ \citep[][their Fig. 3]{Bruens2005}.

\citet{Muller2004} subdivided the MB into four quadrants: two northern ones that ranged from (J2000) $\mathrm{R.A.}\approx 1^h30^m$ to $3^h45^m$ and $\mathrm{decl.}\approx-\ang{73}$ to $-\ang{71}$, and the two southern quadrants at the same RA, but with declinations from $\mathrm{decl.}\approx-\ang{75}$ to $-\ang{73}$. They calculated the spatial power spectrum (SPS) and the spectral correlation function (SCF) for the northern and southern quadrants separately and found that the northern parts appear distinct from the southern parts and from the SMC: the northern parts have an SPS significantly more affected by slower velocity perturbations. They concluded that the MB may not be considered a single contiguous feature but a projection of two kinematically and morphologically distinct structures. Further, they found that the southern region of the MB and the SMC are similar, and that the SPS and SCF of the MB have power-law indices compatible to those found by \citet{Padoan2001} for the LMC. 

\cite{Besla2012} and \cite{Zivick2018} proposed that the MB was caused by a direct collision of SMC and LMC (Besla's+ model 2) where the SMC passed through the LMC about $147\pm33$\,Myr ago \citep{Zivick2018}. 

Moreover, \cite{Zivick2018} favor Model 2 because \citet{Harris2007} observed in situ star formation in the MB and concluded that the stripped material forming the MB was nearly a pure gas. In addition the proper motions of the MB stars track the motion of the SMC toward the LMC. Further evidence for Model 2 comes from \citet{Choi2018}, who find a tidally induced warp in the LMC and conclude that such a warp could only have been created through a direct collision. A two-component radial velocity structure of the \hi\ gas is consistent with this interaction hypothesis, according to which the gas of the LMC forms major parts of the MB \citep{Besla2012}. 

The gas-to-dust ratio differs greatly \citep{RomanDuval2014, Welty2012, Koornneef1982} for the SMC and LMC. Following \cite[][their Sect.\,2.2]{Sasaki2022} we separate the Milky Way and MCS dust emission to constrain the origin of the MB gas seen in \hi\ absorption. For that aim, we adopt their gas-to-dust ratio of $\frac{N_{\mathrm{H}}}{E\mathrm{(B-V)}}=(0.63\pm0.05)\times 10^{22}\left[\frac{\si{\centi\meter}^{-2}}{\mathrm{mag}}\right]$ and multiply that median value with the \hi\ map of the Milky Way column density distribution of HI4PI \citep[][their Fig.\,2]{Sasaki2022}. The difference map between the original \cite{Schlegel1998} (SFD) map and that modeled optical extinction of the Milky Way is displayed in Fig.\,\ref{fig:gastodustratios}. The lowest contour line displayed marks regions two times the Milky Way median value. This contour line encircles the entire MCS system. The highest contour line of five times the median Milky Way value is uniquely reached towards the SMC, both in agreement to \cite{Welty2012} and \cite{Koornneef1982}. The \hi\ absorption lines detected by \cite{Kobulnicky1999} and here (black and white circles respectively) are toward gas volumes with a gas-to-dust ratio compatible with the outskirts of the LMC.

The gas-to-dust ratio of the LMC as reported by \citet{Koornneef1982} based on the International Ultraviolet Explorer data is $N_{\mathrm{H I}}/E(B-V)=(2\pm0.5)\times 10^{22}\,\si{\centi\meter}^{-2}\,\mathrm{mag}^{-1}$. 
\citet{Welty2012} used the Hubble Space Telescope and Far Ultraviolet Spectroscopic Explorer data to estimate $N_{\mathrm{H}}/E(B-V)=(1.5\text{--}1.7)\times 10^{22}\,\si{\centi\meter}^{-2}\,\mathrm{mag}^{-1}$ and $N_{\mathrm{H I}}/E(B-V)=(1.5\text{--}1.6)\times 10^{22} \,\si{\centi\meter}^{-2}\,\mathrm{mag}^{-1}$ as average values for the LMC. Our estimate towards J0332-7249 is closer to the LMC value and less compatible with the average value of $N_{\mathrm{H}}/E(B-V)=(2.3\text{--}3.0)\times 10^{22}\,\si{\centi\meter}^{-2}\,\mathrm{mag}^{-1}$ for the SMC \citep{Welty2012}. Towards J0332-7249 we find for MCS gas-to-dust ratio (Fig.\,\ref{fig:gastodustratios}) $N_{\mathrm{H I}}/E(B-V)=(1.29\pm0.04(\mathrm{stat})\pm0.17(\mathrm{sys}))\cdot 10^{22}\,\si{\centi\meter}^{-2}\,\mathrm{mag}^{-1}$. A value compatible with the hypothesis that the detected MALS \hi\ absorption line is associated with LMC gas.

According to Model 2 of \citet{Besla2012}, the more metal-rich gas from LMC might also have ended up in the MB. This scenario would result in a gradient of increasing metallicity towards the LMC.  The gas-to-dust ratio at 50 locations along the MB and the two measurements from \cite{Koornneef1982} and \cite{Welty2012} are displayed in Fig.~\ref{fig:gastodustratios}. The large drops at both ends of the latitude and longitude ranges correspond to the centers of the LMC and the SMC.  Indeed, as one would expect from Model 2, there is a gradient of decreasing gas-to-dust ratio starting from the SMC \citet[]{SIMBAD2000} ($l=302\fdg 8084$, $b=-44\fdg 3277$) towards the LMC ($l=280\fdg 4652$, $b=-32\fdg 884$, Wenger et al.). For the trend with longitude, the gas-to-dust ratio of the LMC   
extends clearly further than the Magellanic Cloud itself \citep{Welty2012}, which has a radius of about \ang{5}. 

The gas detected in \hi\ absorption exhibits a coherent structure in radial velocity across 4 - 6\,kpc, at least.   Adopting a rather high relative velocity of about 100\,$\mathrm{km\,s^{-1}}$ \citep[][their Fig.\,3]{Bruens2005}, it should have taken about 40 million years for the gas to  move across these distances. Over this time period, the gas should have reached some kind of equilibrium with its environment, allowing us to estimate the pressure within the MB. Towards J0332-7249, observed with the Parkes beam of 16.2\,arcmin in diameter, the average column density is $N_\mathrm{HI}=7.75\times 10^{20}$ \si{\centi\meter}$^{-2}$ . At the adopted distance of 50\,kpc \citep{deGrijs2014} the Parkes beam corresponds to 235~pc or $7.2\times 10^{20}$\,cm. Assuming a depth of the gaseous layer comparable to the width, we can derive an average volume density in the MB of about $n_\mathrm{HI} \simeq 1\,\mathrm{cm^{-3}}$. The line width from Gaussian decomposition yields a Doppler temperature of $T_\mathrm{D}=\SI[separate-uncertainty = true]{857\pm29}{\kelvin}$, an upper limit to the kinetic temperature. The average gas pressure has then an upper limit of $P_\mathrm{K} \sim 900\,\mathrm{K\,cm^{-3}}$. 
\cite{Wang2019} deduced an environmental volume density of about $n_\mathrm{MW} \simeq 2\,-\,3\times 10^{-4}\,\mathrm{cm^{-3}}$ (their Fig.\,1 bottom panel) to account for the observed density and radial--velocity structure of the MCS. To enable a pressure equilibrium between this ambient coronal plasma and the \hi\ clouds presented here, the plasma needs to be very hot, about $T_\mathrm{Halo} \sim 2\times10^6$\,K. Such a hot thin plasma gas has been observed by \cite{Sasaki2022}. Notably, the values ($P = n\,\mathrm{k}T_\mathrm{kin}(\mathrm{HI})$, $n_\mathrm{H}$, $T_\mathrm{MW}$) discussed above are consistent with three independent studies on the morphology of the MCS, the \hi\ data of \cite{Kobulnicky1999} and the one presented here as well as soft X-ray studies. 

\section{Conclusions}

We report the detection of an \hi\ 21-cm absorption line associated with the MB  towards the MALS sightline J0332-7249. 
In combination with the two \hi\ absorption lines detected by \cite{Kobulnicky1999}, this enables us to probe the average gas volumic density of the MB over 4-6\,kpc. Despite the kpc-scale separation, all three absorption lines show that the cold dense gas traced by \hi\ absorption is at the low radial velocity part of the corresponding \hi\ emission profile, a common feature of the gas associated with the MB. All \hi\ 21-cm emission profiles reveal a two-component structure in velocity space. The absorbing gas consistently coincides with a higher peaked, narrower \hi\ emission line component at lower radial velocities, implying cold and dense gas. This also suggests that the gas is decelerated by ram--pressure from the Milky Way's halo plasma across linear scales of several kpc. There might also exist gas exchanges between the two MCS galaxies. The derived spin temperature of $T_\mathrm{S} = 104\,\pm\,20$\,K is consistent with the range of temperatures derived by \cite{Kobulnicky1999}.  
The similarity between the MB gas-to-dust ratio and that of the LMC, rather than the SMC, supports an origin of the MB gas in the LMC. The SMC's gas-to-dust ratio is far higher, implying that the MB gas was extracted from the LMC. According to \cite{Zivick2018} and \cite{Besla2012}, the direct collision of the SMC and LMC is currently favored. Our 21-cm study supports the direct LMC-SMC collision scenario.

\begin{acknowledgements}
We thank an anonymous referee for very constructive critics and suggestions significantly improving the presentation.
The MeerKAT telescope is operated by the South African Radio Astronomy Observatory, which is a facility of the National Research Foundation, an agency of the Department of Science and Innovation. 
The MeerKAT data were processed using the MALS computing facility at IUCAA (\url{https://mals.iucaa.in/releases}).
%
%
NG acknowledges NRAO for generous financial support for the sabbatical visit at Socorro during which a part of this work was done. 
The National Radio Astronomy Observatory is a facility of the National Science Foundation operated under cooperative agreement by Associated Universities, Inc.  
%
\end{acknowledgements}


%
%

\bibliographystyle{aa_url.bst} 


\begin{appendix}

\section{Gaussian decomposition of emission and absorption lines}
\label{sec:gauss}

\begin{table}[!htbp]
    \caption{\hi\ emission lines, Gaussian fits}
     
	\centering
	\begin{tabular}{ccc}
       \hline\hline
    & MW Line & MB Line \\
    \hline
     Peak - $T_A$ [\si{\kelvin}] &  & \\
     1st & 11.1 &  2.2 \\
     2nd & 4.4 & 1.7\\
     3rd & 4.4 & 1.3\\
     4th & 3.7 & 0.7\\
		\hline
    Center [\si{\kilo\meter\second}$^{-1}$] & & \\
     1st & 2.4 & 239.8\\
     2nd & 7.3 & 219.1\\
     3rd & -2.2 & 219.6\\
     4th & -1.6 & 228.3\\
     \hline
     Width - $\sigma$ [\si{\kilo\meter\second}$^{-1}$] &  & \\
      1st & 2.2 & 18.4 \\
     2st & 2.8 & 10.3\\
     3nd & 3.8 & 4.9 \\
     4th & 9.6 & 2.3 \\
		\hline
	\end{tabular}
	\label{tab:emissionfits}
    \tablefoot{HI4PI Gaussian fits results (see Fig.\ref{fig:Magellanic3}) towards J0332-7249 
     by \citet[][]{KalberlaHaud2018}. The uncertainties in peak brightness $T_\mathrm{A}$ and line width $\sigma$ correspond to the inherent calibration uncertainties of about 5\% \citep{HI4PI}, in $\Delta v_\mathrm{LSR} \leq 1.28\,\mathrm{km\,s^{-1}}$.}
\end{table}

%
%
%
%
%
\begin{figure}[!htbp]
	\centering
	\includegraphics[width=\hsize]{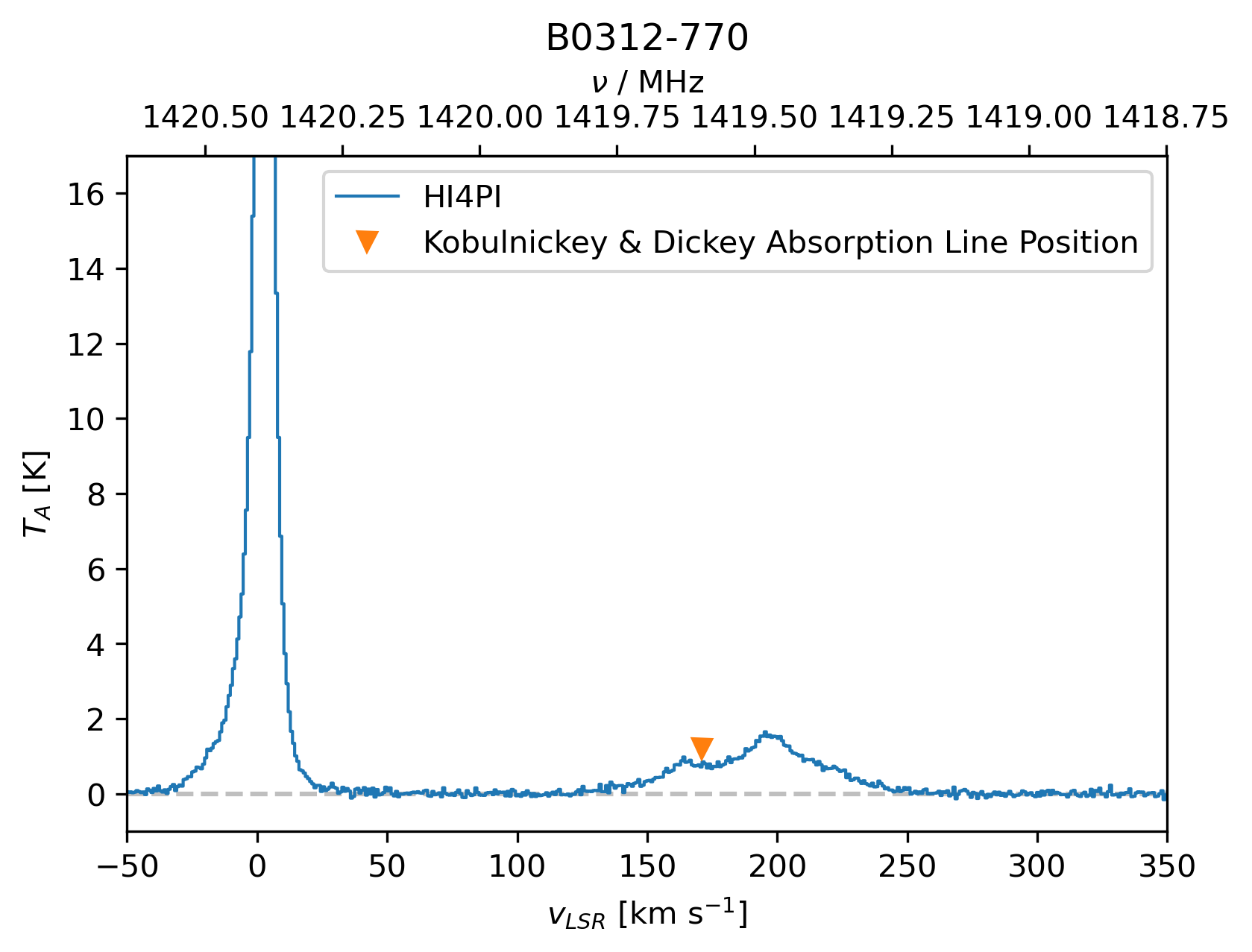}
    \includegraphics[width=\hsize]{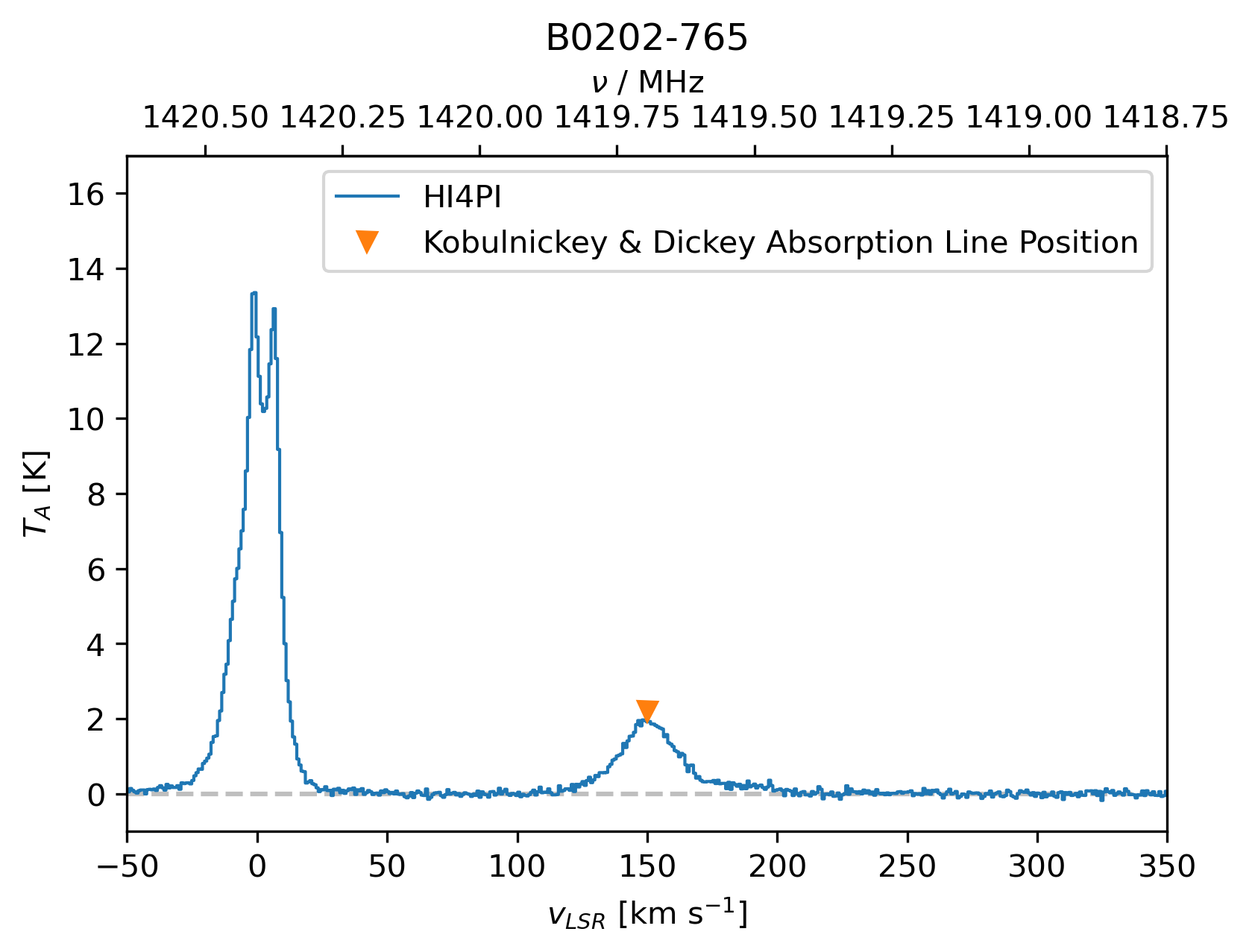}
	\caption{HI4PI emission spectra from the \citet{HI4PImapsite} towards B0312-770 and B0202-765 at which \citet{Kobulnicky1999} detected absorption lines. The velocity component of the absorption line with the highest column density is marked with an orange triangle.
    }
	\label{fig:Magellanic4}
\end{figure}


\begin{table}[!htbp]
    \caption{Details of Gaussian fits to absorption lines towards J0332-7249 applying a single Gaussian fit.}
	\centering
	\begin{tabular}{ccc}
		\hline\hline
    & MW Line & MB Line \\
    \hline
     Peak [$10^{-3}$\si{\jansky}/beam$^{-1}$] & $-7.3\pm1.1$ &$-8.6\pm1.7$ \\
	
     \hline
     Center [\si{\kilo\meter\second}$^{-1}$] &$3.27\pm0.76$ & $224.59\pm0.97$\\

     \hline
     $\sigma$ [\si{\kilo\meter\second}$^{-1}$] & $4.13\pm0.71$ & $2.66\pm0.49$\\
    
		\hline
	\end{tabular}
	\label{tab:absorptionfits}
\end{table}

\end{appendix}





   
  



\end{document}